\documentclass[%
 aip,
jmp,
 amsmath,amssymb,
 reprint,%
]{revtex4-1}

\usepackage{graphicx}% Include figure files
\usepackage{dcolumn}% Align table columns on decimal point
\usepackage{bm}% bold math
\usepackage[utf8]{inputenc}
\usepackage[T1]{fontenc}
\usepackage{mathptmx}

\newcommand{\msr}{$\mu$SR\xspace}
\newcommand{\IMSS}{Muon Science Laboratory and Condensed Matter Research Center, Institute of Materials Structure Science, High Energy Accelerator Research Organization, Tsukuba, Ibaraki 305-0801, Japan}
\newcommand{\Sokendai}{Department of Materials Structure Science, The Graduate University for Advanced Studies (Sokendai),  Tsukuba, Ibaraki 305-0801, Japan}
\newcommand{\MSLTitech}{Laboratory for Materials and Structures, Institute of Innovative Research, Tokyo Institute of Technology, Yokohama, Kanagawa 226-8503, Japan}
\newcommand{\MCES}{Materials Research Center for Element Strategy, Tokyo Institute of Technology (MCES), Yokohama, Kanagawa 226-8503, Japan}

\usepackage{xspace}
\usepackage{color}
\usepackage{ulem}

\begin{document}

\title{Electronic structure of interstitial hydrogen in In-Ga-Zn-O semiconductor simulated by muon}%
%\thanks{Footnote to title of article.}

\author{K. M. Kojima} 
\affiliation{\IMSS}\affiliation{\Sokendai}
\author{M. Hiraishi}
\affiliation{\IMSS}
\author{H. Okabe}
\affiliation{\IMSS}
\author{A. Koda}
\affiliation{\IMSS}\affiliation{\Sokendai}
\author{R. Kadono}\altaffiliation[Corresponding author: ]{ryosuke.kadono@kek.jp}\affiliation{\IMSS}\affiliation{\Sokendai}
\author{K. Ide}
\affiliation{\MSLTitech}
\author{S. Matsuishi}
\affiliation{\MCES}
\author{H. Kumomi}
\affiliation{\MCES}
\author{T. Kamiya}
\affiliation{\MSLTitech}\affiliation{\MCES}
\author{H. Hosono}
\affiliation{\MSLTitech}\affiliation{\MCES}

\date{\today}%

\begin{abstract}

We report on the local electronic structure of interstitial muon (Mu) as pseudo-hydrogen in In-Ga-Zn oxide (IGZO) semiconductor studied by muon spin rotation/relaxation (\msr) experiment.  In polycrystalline (c-) IGZO, it is inferred that Mu is in a diamagnetic state,  where the \msr\ time spectra under zero external field is perfectly described by the Gaussian Kubo-Toyabe relaxation function with the linewidth $\Delta$ serving as a sensitive measure for the random local fields from In/Ga nuclear magnetic moments. The magnitude of $\Delta$ combined with the density functional theory calculations for H (to mimic Mu) suggests that Mu occupies Zn-O bond-center site (Mu$_{\rm BC}$) similar to the case in crystalline ZnO.  This implies that the diamagnetic state in c-IGZO corresponds to Mu$_{\rm BC}^+$, thus serving as an electron donor.  In amorphous (a-) IGZO, the local Mu structure in as-deposited films is nearly identical with that in c-IGZO, suggesting Mu$_{\rm BC}^+$ for the electronic state. In contrast, 
the diamagnetic signal in heavily hydrogenated a-IGZO films exhibits the Lorentzian Kubo-Toyabe relaxation, implying that Mu accompanies more inhomogeneous distribution of the neighboring nuclear spins that may involve Mu$^-$H$^-$-complex state in an oxygen vacancy.  
\end{abstract}

\maketitle

%\section{Introduction}
Amorphous oxide semiconductors including InGaZn oxide (IGZO) are rapidly gaining success as channel materials in transparent thin-film transistors (TFTs), used as the main large-area semiconductor in display applications such as liquid-crystal displays and organic light-emitting diode displays.\cite{Hosono:06,Kamiya:10}  These compounds exhibit relatively high electron mobility due to the $s$-like character of the conduction band edge states which are insensitive to disorder.\cite{Hosono:06,Kamiya:10,Robertson:12}
However, the operating condition for TFT that they are always illuminated by the backlight often leads to a severe problem for a-IGZO known as negative bias illumination stress (NBIS),\cite{Ghaffarzadeh:10,Lee:10,Chowdhury:10,Park:12,Jeon:12,Jeong:13} whose microscopic mechanism is still under debate. 
 The photoexcitation of electrons from in-gap defect levels lying right above the valence-band top to the conduction band results in a persistent photoconductivity, causing the shift of operating bias voltage in the TFTs. Evidence for the corresponding in-gap states has been reported by studies using hard x-ray photoemission spectroscopy.\cite{Nomura:11,Nomura:19}

In a-IGZO, hydrogen (H) has been known as an important agent for improving various electronic properties. For example, the carrier mobility and the $S$-value (sharpness of the TFT switching as a function of Gate-Source voltage) are improved under water vapor annealing, where H is presumed to play a central role.\cite{KamiyaJDT09}  Recent infrared transmission spectroscopy has characterized the deep state at the vicinity of the valence band maximum, where the absorption mode frequencies are consistent with those estimated for a model structure of oxygen vacancy filled by two hydride ions (2H$^-$ at V$_{\rm O}$) coupled to the surrounding cations.\cite{BangAPL17,Li:17}  In addition, H treatment of IGZO is known to cause an $n$-type conductivity. Thus, understanding the microscopic mechanisms of these processes would be of crucial importance in controlling the electronic properties of IGZO.

It is well established that a positive muon ($\mu^+$) implanted into matter can be regarded as a light radioisotope of proton in the sense that the local structure of a muon-electron system is virtually equivalent with that of H, except for a small correction ($\simeq0.4$\%) due to the difference in the reduced {\sl electron} mass. While the light mass of muon ($\simeq m_{\rm p}/9$, with $m_{\rm p}$ being the proton mass) often leads to the isotope effect which is particularly distinctive in diffusion at low temperatures where quantum tunneling process becomes dominant, muon also simulates proton/H diffusion via classical over-barrier jump at ambient temperature.  Thus, muon in matter can be regarded as a pseudo-hydrogen. We proposed the designation ``muogen" (Mu) as the appropriate elemental name,\cite{Okabe:18} because the term ``muonium" exclusively refers to the neutral bound state of $\mu^+$ and $e^-$ (Mu$^0$) analogous to  H$^0$. It would be also worth stressed that the information derived from implanted Mu corresponds to that for H in its dilute limit, considering the small number of implanted muons ($<10^5$ cm$^{-3}$s$^{-1}$) and their short lifetime ($\simeq2.2$ $\mu$s) for the beta-decay into positrons and neutrinos (i.e., no accumulation of Mu upon muon irradiation).

In this Letter, we discuss the local electronic structure of H in IGZO simulated by Mu.
The Mu site is determined for the polycrystalline specimen (c-IGZO) by the magnitude of random local fields from nuclear magnetic moments [described by the second moment of internal field distribution, $\Delta^2$ (or its rms value $\Delta$)], where the density functional theory (DFT) calculation is used to narrow down the candidate sites.  Subsequently, the local environment of Mu in amorphous sample (a-IGZO) is estimated on the basis of result obtained in c-IGZO.  The possibility of Mu$^-$H$^-$-V$_{\rm O}$ complex defects in heavily hydrogenated a-IGZO is also discussed.

%\section{Experimental}
%\subsection{IGZO samples}
Polycrystalline specimen of IGZO was prepared by solid state reaction of In$_2$O$_3$, Ga$_2$O$_3$ and
ZnO powders at 1450 $^\circ$C in a Al$_2$O$_3$ crucible under atmospheric condition. The obtained specimen in a sintered pellet, which was originally used as the target for growing IGZO thin films by pulsed laser deposition, had $\approx$85\% of the density for the ideal crystal (which was considerably smaller than that of commercially available sputter target $\sim$98\%), probably due to the relatively low sintering temperature.  The pellet was then sliced into slabs with thickness of $\sim$0.5 mm, and subjected to O$_2$ annealing at 1000 $^\circ$C. The oxidization was monitored by the color of the pellet that turned from gray to white after annealing. 

Amorphous IGZO films were prepared by RF magnetron sputtering onto oxidized Si substrate without intentional substrate heating.  Deposition condition was optimized for making reasonable IGZO-TFT, where the film had a carrier concentration of $\sim$10$^{14}$\,cm$^{-3}$. 
A part of the as-deposited films was then subjected to H-plasma treatment with the condition reported elsewhere,\cite{TangSSST17} leading to the increase of electron concentration to $\sim$10$^{19}$\,cm$^{-3}$. The thickness of the film was determined to be 200$\pm$10 nm by the X-ray reflectivity measurement. Hereafter, we refer to the as-deposited and H-treated samples as a-IGZO and a-IGZO:H, respectively.

%\subsection{Muon experiment} 
The local electronic structure of Mu can be assessed by the muon spin rotation/relaxation (\msr) technique, where the magnitude and distribution of the local magnetic field ($B_{\rm loc}$) at the Mu site is monitored by the time evolution of muon spin polarization $P_\mu(t)$. In the paramagnetic state (Mu$^0$) where an unpaired electron is bound to muon, $B_{\rm loc}$ is predominantly determined by the muon-electron hyperfine field ($A_\mu$). The corresponding \msr\ time spectrum under a transverse field (TF, $B$ perpendicular to the initial muon polarization) exhibits a multiplet structure [$P_\mu(t)=\sum_{i=1}^3a_i\cos\omega_it$, where the amplitude $a_i$ and frequency $\omega_i$ are determined by $A_\mu$ and $B$].\cite{Holzschuh:82} It is also important to note that these multiplet signals are subject to strong broadening when the unpaired electron is coupled with surrounding nuclear spins (called nuclear-hyperfine interaction).\cite{Kiefl:86}

In the case of the diamagnetic state  (Mu$^+$ or Mu$^-$, $A_\mu=0$),  $B_{\rm loc}$ is determined by the random local fields exerted from nearby nuclear magnetic moments. While each Mu exhibits the Larmor precession [$P_\mu(t)=\cos\omega_\mu t$] with a frequency $\omega_\mu=\gamma_\mu B_{\rm loc}$  ($\gamma_\mu=2\pi\times 135.53$ MHz/T being the muon gyromagnetic ratio),  $P_\mu(t)$ as an ensemble of these precession signals is given by an inverse Fourier transform of the probability function $P(\omega_\mu)$. In particular, when a number of the nearest neighboring (nn) nuclei are present at nearly equal distance from Mu, the Gaussian distribution is a good approximation for $P(\omega_\mu)$, and the corresponding $P_\mu(t)$ under a zero external field ($B=0$) in polycrystalline specimen is given by the Gaussian Kubo-Toyabe relaxation function\cite{HayanoPRB79}
\begin{equation}
P_\mu(t)= \frac{1}{3} + \frac{2}{3}[1-(\Delta t)^2]\cdot\exp\left[-\frac{1}{2}(\Delta t)^2\right]=G_{\rm KT}(t)\label{eq:GKT},
\end{equation}
where $\Delta$ corresponds to the second moment of $P(\omega_\mu)$.  More specifically, $\Delta$ is evaluated as a sum of contributions from the $n$-th kind of nuclear magnetic moments ($n = 1,2,3,4,$ and 5 for $^{67}$Zn, $^{69}$Ga, $^{71}$Ga, $^{113}$In, and $^{115}$In),
\begin{eqnarray}
\Delta^2&\simeq&\gamma_\mu^2\sum_{n}f_n\sum_j\sum_{\alpha=x,y}\sum_{\beta=x,y,z}\gamma_n^2({\bf \hat{A}}_{nj}{\bf I}_n)^2 \label{dlts}\\
{\bf \hat{A}}_{nj}&=&A^{\alpha\beta}_{nj}=(3\alpha_j\beta_j-\delta_{\alpha\beta}r_{nj}^2)/r_{nj}^5\quad(\alpha, \beta=x,y,z)\nonumber
\end{eqnarray}
with ${\bf r}_{nj}=(x_j,y_j,z_j)$ being the position vector of the $j$-th nucleus from Mu,  ${\bm \mu}_n=\gamma_n{\bf I}_n$ the nuclear magnetic moment with $f_n$ being their natural abundance.   Because all the nuclei in IGZO have spin $I_n\ge1$, ${\bm \mu}_n$ is subject to electric quadrupolar interaction with the electric field gradient generated by the point charge of the diamagnetic Mu. This leads to the reduction of effective ${\bm \mu}_n$ to the value parallel with ${\bf r}_{nj}$ (by a factor $\sqrt{2/3}$ in the classical limit).\cite{HayanoPRB79}
In any case, the typical magnitude of $\Delta^{-1}$ falls in the time range of $\mu$SR ($10^{-5}$--$10^{-6}$ s), over which the nuclear moments are quasistatic with random orientation. Since $f_n$ and ${\bm \mu}_n$ are known for the host compounds, one can estimate the Mu location from the magnitude of $\Delta$ which strongly depends on $r_{nj}$.  

While the nuclear dipolar fields are quasistatic, the diffusive motion of Mu induces dynamical modulation of $G_{\rm KT}(t)$. In particular, the 1/3 component in Eq.~(\ref{eq:GKT}) is sensitive to slow Mu diffusion, as it disappears when the Mu hopping rate $\nu$ becomes comparable with $\Delta$. The lineshape of $G_{\rm KT}(t)$ becomes exponential for the fast Mu diffusion ($\nu\gg\Delta$), which is approximately given by $G_{\rm KT}(t)\simeq\exp(-\Delta^2t/\nu$).  Since the spin relaxation induced by the dynamical modulation is virtually unaffected by a longitudinal field (LF, $B$ parallel with initial muon polarization) in contrast to the quasistatic limit where $G_{\rm KT}(t)\simeq1$ for $B\gg\Delta/\gamma_\mu$,  the presence of such a dynamical effect is experimentally examined by the response of $P_\mu(t)$ against LF.\cite{HayanoPRB79}

\begin{figure}[t]
\begin{center}
  \includegraphics[width=0.95\columnwidth]{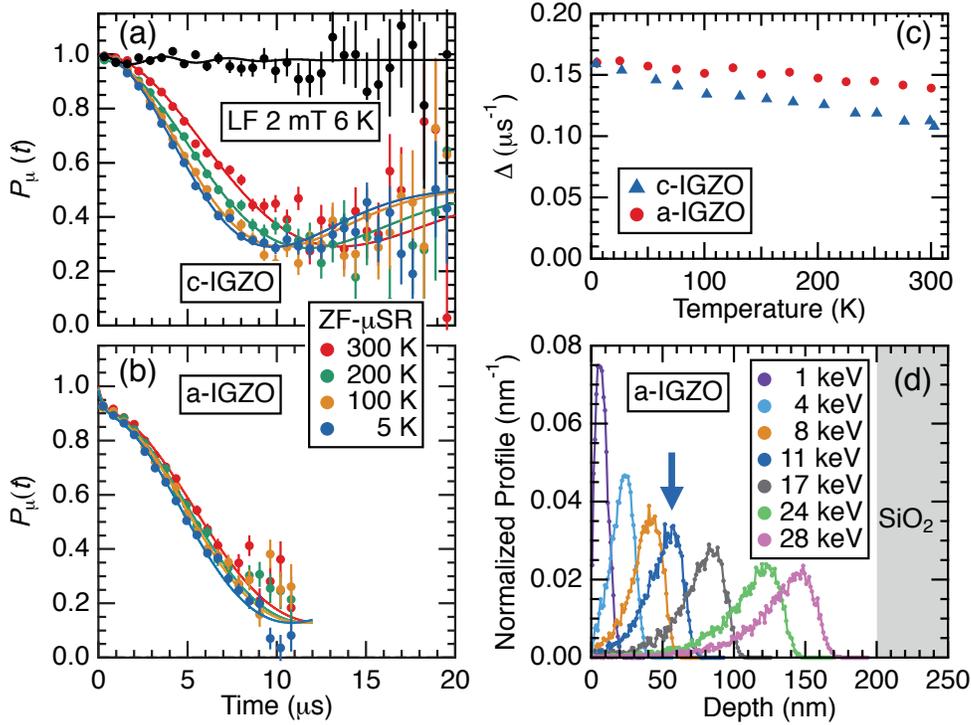}
\end{center}
\caption{(Color online) 
 \msr\ spectra observed in (a) c-IGZO and in (b) as-deposited a-IGZO specimens at various temperatures under zero (ZF-) or a longitudinal field (LF), where solid curves represent least-square fit by the Gaussian Kubo-Toyabe function.
(c) Temperature dependence of the Gaussian linewidth $\Delta$ in c-IGZO and as-deposited a-IGZO.  (d) Muon depth profiles in a-IGZO film ($\sim$200 nm thick, grown on SiO$_2$ substrate) for various incident kinetic energy $E_\mu$ simulated by TRIM.SP (see text), where the present data were obtained with $E_\mu=11$ keV.}

\label{fig:IGZOspectra}
\end{figure}

With these characteristic features of \msr\ in mind, conventional \msr\ measurements on c-IGZO specimen were performed using the ARTEMIS spectrometer\cite{KojimaJPSConf17} installed in the S1 area of Material and Life Science Experimental Facility, J-PARC where a nearly 100\% spin-polarized pulsed beam of high-flux ``surface muon"  (repetition rate of 25 Hz, pulse width $\sim$100 ns, with a beam energy $E_\mu\simeq4$ MeV) was available. The time evolution of $P_\mu(t)$ was monitored by the decay-positron asymmetry [$A(t)$] measured by two sets of scintillation telescopes placed forward/backward position against sample position, where  $P_\mu(t)=A(t)/A_0$ with typical instrumental asymmetry $A_0\simeq0.2$. We employed a ``fly-past" chamber set-up for minimizing background positrons from muons which missed the sample. Meanwhile, \msr\ measurements on a-IGZO and a-IGZO:H films were conducted using Low Energy Muon beam ($E_\mu\le 30$ keV) provided at Paul Scherrer Institute, Switzerland. The muon stopping profiles in those films (5.97 g/cm$^3$) were optimized by a Monte Carlo simulation using TRIM.SP (see Fig.~\ref{fig:IGZOspectra}d).\cite{trim:91}  The details of calculating $\Delta$ using Eq.~(\ref{dlts}) for the candidate Mu sites are found elsewhere.\cite{KojimaPRB04}

Typical \msr\ time spectra observed at various temperatures in c-IGZO specimen are shown in Fig.~\ref{fig:IGZOspectra}a, where the lineshape clearly exhibits a Gaussian damping that is accompanied by the recovery of polarization at later times ($t\agt10$ $\mu$s). 
These spectra are perfectly reproduced by Eq.~(\ref{eq:GKT}), indicating that i) Mu is entirely in the diamagnetic state, and that ii) the relaxation is induced by the random local fields exerted from nuclear magnetic moments. The quasistatic character of the local field is further corroborated by the response of spectra to LF, where the relaxation is completely suppressed by applying a small field that exceeds $\Delta/\gamma_\mu\simeq0.2$ mT  (see the \msr\ spectrum with LF = 2 mT at 6 K).\cite{HayanoPRB79} The magnitude of $\Delta$ deduced by curve-fit using Eq.~(\ref{eq:GKT}) is shown in Fig.~\ref{fig:IGZOspectra}c.

%%%%%%%%%%%%%%%%%%%%%%%%%%%%%%%%%%%%%%%%%%%%%%%%%%%%
%%%%%%%%%%%%%%%%%%%%%%%%%%%%%%%%%%%%%%%%%%%%%%%%%%%%%%%%%

%Interestingly, $\Delta$ exhibits gradual decrease with temperature 

At this stage, it would be interesting to compare these results with that for the a-IGZO films shown in Fig.~\ref{fig:IGZOspectra}b.  Despite the relatively narrow time range of observation ($\sim$11 $\mu$s, limited by background noise), the Gaussian damping consistent with Eq.~(\ref{eq:GKT}) is observed with the relaxation rate similar to that for c-IGZO.  This suggests that the local environment of Mu in a-IGZO is nearly identical with that in c-IGZO, as long as the configuration of the nn atoms (nuclei) around Mu is concerned.
The solid curves in Fig.~\ref{fig:IGZOspectra}b show the result of curve-fit using Eq.~(\ref{eq:GKT}) [where the small reduction of $P_\mu(t)$ near $t=0$ due to the muon backscattering was considered by additional component in the curve-fit with fast exponential damping], yielding $\Delta$ at various temperatures as shown in Fig.~\ref{fig:IGZOspectra}c. While $\Delta$ shows nearly perfect agreement with that in c-IGZO ($\Delta\simeq0.16$ $\mu$s$^{-1}$) below $\sim$30 K, it exhibits a less pronounced reduction at higher temperatures. 

It is important to note that the \msr\ lineshape remains to be perfectly reproduced by Eq.~(\ref{eq:GKT}) regardless of the variation in $\Delta$. In particular, the recovery of $P_\mu(t)$ for $t\ge10$ $\mu$s in c-IGZO clearly demonstrates absence of Mu diffusion up to 300 K (i.e., Mu hopping rate is much smaller than $\Delta\sim10^{5}$ s$^{-1}$). Thus, the temperature dependence of $\Delta$ cannot be attributed to the long-range Mu diffusion.  This leads to a scenario of multiple Mu sites, where the lowest energy site is separated from other metastable sites by potential barriers. It is further presumed that, while the initial occupation of these sites upon muon implantation is determined by the respective densities, their actual population at a given temperature is modulated by the principle of detailed balance.

In order to narrow down the candidate Mu sites where $\Delta$ calculated by Eq.~(\ref{dlts}) is consistent with the values in Fig.~\ref{fig:IGZOspectra}c, we performed the DFT calculation for H (to mimic Mu) using VASP code package.\cite{VASP} We presumed that the local electronic structures as well as energetics for H would be nearly identical with those for Mu because of the small difference in the effective electron mass ($\sim$0.4\%) and zero-point energy ($\sim$0.1-0.2 eV). A supercell of 2$\times$1$\times$1 c-IGZO lattice units was prepared to host one H atom, and the electron energy was minimized in terms of the H position. Structural relaxation was also incorporated for a few arbitrary starting H positions. The formation energy $E_f$ for the converged structure was calculated by comparing the total electron energy of the relaxed final state with that for the pristine c-IGZO lattice plus atomic H.  The obtained electron density for the final state was in the order of $n\sim$10$^{20}$cm$^{-3}$, yielding the corresponding Fermi level $E_{\rm F}$ within the conduction band situated approximately 0.5\,eV above the conduction band minimum.

%\vspace{0.5cm}
%%%%%%%%%%%%%%%%%%%%%%%%%%%%%%%%%%%%%%%%%%%%%%%%%%%%
\begin{figure}[t]
\begin{center}
  \includegraphics[width=0.9\columnwidth]{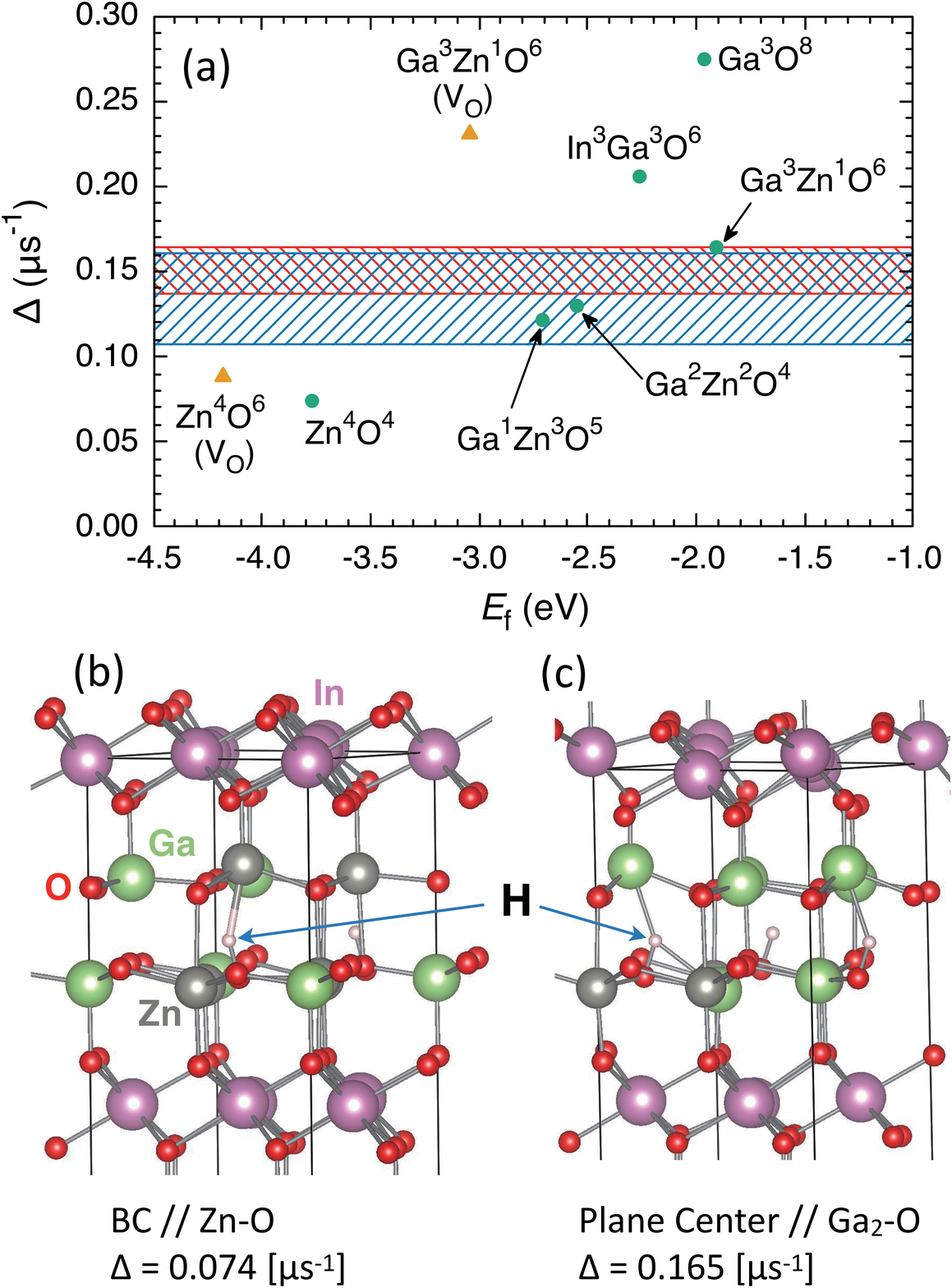}
\end{center}
\caption{(Color online) 
(a) Summary of $E_f$ vs.~$\Delta$ calculated for the Mu candidate sites, where the label $A^lB^mC^n$ indicates the element name and number of atoms located within 0.3 nm from Mu sites. The hatched band areas indicate the range of $\Delta$ for c-IGZO and a-IGZO shown in Fig.~\ref{fig:IGZOspectra}c.  Some examples of local atomic configurations are shown for the (b) Zn$^4$O$^4$ site (Zn-H distance $=0.221$ nm, O-H distance $=0.098$ nm) and (c) Ga$^3$Zn$^1$O$^6$ site (Ga-H  $=0.208$ nm,  O-H  $=0.100$ nm). }
\label{fig:DFT}
\end{figure}
%%%%%%%%%%%%%%%%%%%%%%%%%%%n%%%%%%%%%%%%%%%%%%%%%%%%%%%%%%, or (d) those at the oxygen vacancy positions (V$_{\rm O}$)

The calculated $E_f$ and $\Delta$ are shown in Fig.~\ref{fig:DFT} for the various H sites.  It must be noted that the occupation of Ga and Zn atoms is random in their crystallographic sites, allowing a variety of Ga/Zn configurations for the nn sites around the H atom. It is inferred that $E_f$ tends to be reduced by the increase of Zn occupancy for the nn sites against Ga, showing a minimum at the Zn-O bond-centered position (BC$_\parallel$). A similar result is obtained for H in the oxygen vacancy (V$_{\rm O}$, see Fig.~\ref{fig:DFT}c)
Moreover, Ga ions are found to be expelled from the H position in the process of structural relaxation. It is also inferred that $E_f$ is even higher for H placed near In ions in comparison to  other cases.  These results suggest that H prefers the ZnO/GaO layers to InO$_2$ layers, and that the H sites near Zn are more stable in the mixed Zn/Ga coordinations. 

The reasonable agreement between the experimentally deduced $\Delta$ in c-IGZO at ambient temperature [$\simeq0.110(1)$ $\mu$s$^{-1}$] and those for the interstitial sites in the Ga/ZnO$_2$ layers ($\simeq 0.122$  and 0.130 $\mu$s$^{-1}$ for Ga$^1$Zn$^3$ and Ga$^2$Zn$^2$ sites, respectively) indicates that Mu/H occupies the Zn-rich sites. This also suggests that, apart from the difference in the relative orientation of ZnO$_4$ tetrahedra between IGZO and ZnO, the local electronic structure of Mu/H is similar to that in ZnO where Mu is presumed to occupy BC$_\parallel$ sites.\cite{CGvWalle:01,ShimomuraPRL02,Kilic:02}  Considering that the electronic structure of Mu in ZnO (corresponding to that in the dilute limit of the interstitial H) is that of a typical shallow-donor state with small $A_\mu$ ($\sim10^{-4}$ of the vacuum value), it is speculated that the electronic level associated with the Mu$_{\rm BC}$ state in c-IGZO is slightly shifted upward from that in ZnO to resonate with the conduction band, yielding Mu$_{\rm BC}^+$ as the most stable electronic state. This is also consistent with the observation that H-plasma treatment of a-IGZO leads to strong $n$-type doping at lower H doses, suggesting H$_{\rm BC}$ for the corresponding H center. 

Since the contribution of Zn nuclei to $\Delta$ is negligible (as the natural abundance of $^{67}$Zn is as small as 4.1\%),  Eq.~(\ref{dlts}) indicates that $\Delta^2$ is nearly proportional to the number of nn Ga ions.  Thus, $\Delta$ should exhibit a positive correlation with $E_f$, which is readily observed in Fig.~\ref{fig:DFT}a. The gradual decrease of $\Delta$ with increasing temperature observed in c/a-IGZO (see Fig.~\ref{fig:IGZOspectra}c) is understood as the re-distribution of Mu site occupancy from one determined statistically to the other which is closer to the thermal equilibrium. The tendency of larger $\Delta$ in a-IGZO than that in c-IGZO may be attributed to the lower Mu mobility due to random potentials in the amorphous structure. Thus, we conclude that local electronic structure of Mu is essentially identical between c-IGZO and a-IGZO except minor difference in the nn Ga/Zn configuration.

%%%%%%%%%%%%%%%%%%%%%%%%%%%%%%%%%%%%%%%%%%%%%%%%%%%%
\begin{figure}[t]
\begin{center}
  \includegraphics[width=\columnwidth]{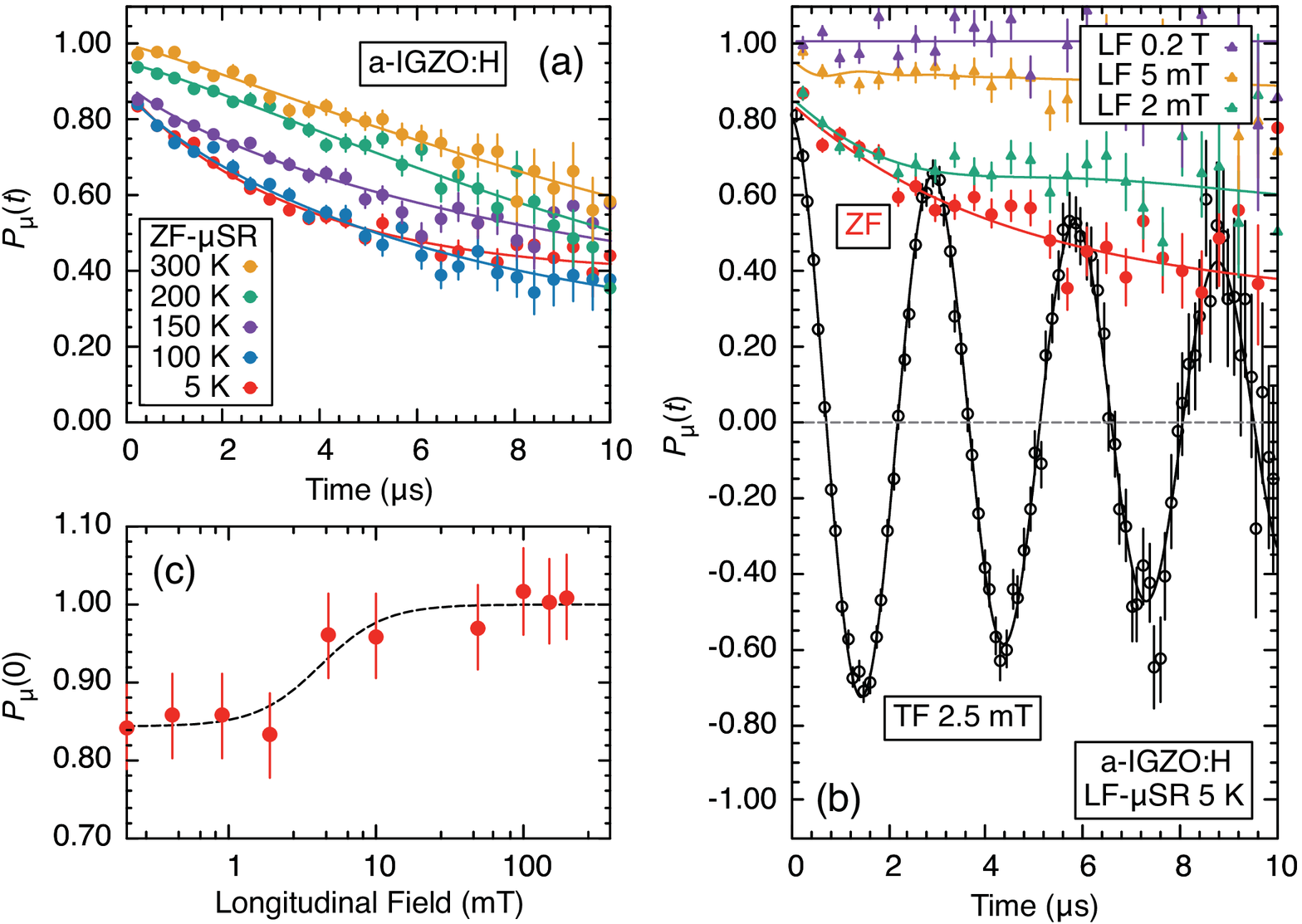}
\end{center}
\caption{(Color online) (a), (b) \msr\ spectra measured under various conditions in H-treated a-IGZO. (c) Magnetic-field dependence of the initial muon polarization, where the horizontal axis is in a logalithmic scale.}
\label{fig:IGZOHspectra}
\end{figure}
%%%%%%%%%%%%%%%%%%%%%%%%%%%%%%%%%%%%%%%%%%%%%%%%%%%%%%%%%

In contrast, the \msr\ spectra in a-IGZO:H films exhibit qualitatively different behavior from those in a-IGZO films. As shown in Fig.~\ref{fig:IGZOHspectra}a, a part of the initial muon polarization [$P_\mu(0)$] is missing at lower temperatures, and the remaining $P_\mu(t)$ exhibits a slow exponential relaxation corresponding to the earlier part ($t\le\Delta$) of the Lorentzian Kubo-Toyabe function [where $(\Delta t)^2$ in Eq.(\ref{eq:GKT}) is replaced by $\Delta t$].\cite{Crook:97} The response of the latter component to a TF  ($B =2.5$ mT, see Fig.~\ref{fig:IGZOHspectra}b) showing spin precession with a frequency $\gamma_\mu B=0.339$ MHz indicates that the component is entirely attributed to a diamagnetic state (Mu$_{\rm d}$). Furthermore, the absence of fast dynamical modulation is confirmed by the spectra observed under LF in which the relaxation is suppressed by $B\simeq2$ mT.  These features strongly suggest that, while the relaxation of Mu$_{\rm d}$ is induced by the quasistatic local fields from nuclear magnetic moments, they are characterized by the inhomogeneous distribution of $r_{nj}$ as well as $n$.  This is qualitatively in line with the presumption that the local structure around Mu$_{\rm d}$ involves another H atom, which also recalls a model of 2H$^-$ in V$_{\rm O}$ proposed as a candidate of the deep center in a-IGZO.\cite{BangAPL17,Li:17} A curve fit of the spectra in Fig.~\ref{fig:IGZOHspectra}b by the Lorentzian Kubo-Toyabe function yields $\Delta=0.148(17)$ $\mu$s$^{-1}$, which is considerably greater than 0.088 $\mu$s$^{-1}$ expected for the isolated Mu in the V$_{\rm O}$ site (i.e.,  Zn$^4$O$^6$ site in Fig.~\ref{fig:DFT}a).  While the nearby H is expected to induce sinusoidal components specific to the two-spin system with a characteristic angular frequency $\omega_d=2\gamma_\mu\gamma_{\rm p}/r_{\rm \mu p}^3$ for $P_\mu(t)$ (where $\gamma_{\rm p}$ is the gyromagnetic ratio of proton, and $r_{\rm \mu p}$ is the distance between Mu and H),\cite{Kadono:08} the strong dependence of $\omega_d$ to $r_{\rm \mu p}$ would lead to the smearing of these components to the Lorentzian-like lineshape due to the variation of $r_{\rm \mu p}$ in the amorphous IGZO matrix.   It is tempting to speculate that Mu in a-IGZO:H tends to find unpaired H around V$_{\rm O}$, forming Mu$^-$-H$^-$ complex.
 
Meanwhile, the response of $P_\mu(0)$ to LF at higher fields ($B>2$ mT, see Fig.~\ref{fig:IGZOHspectra}c) indicates that the missing polarization corresponds to a quasistatic paramagnetic state (Mu$_{\rm p}$).  The complete loss of the initial Mu$_{\rm p}$ polarization is readily understood by considering the nuclear-hyperfine (NHF) interaction with Ga nuclei that leads to fast relaxation of the spin-triplet state.\cite{Kiefl:86,Beck:75} As seen in Fig.~\ref{fig:IGZOHspectra}c, the recovery of $P_\mu(0)$ with increasing $B$ is in single-step over a narrow field range of $\alt$10 mT, implying that the magnitude of $A_\mu$ is less than that of the NHF interaction.  In addition, the recovery of $P_\mu(0)$ to $\sim$1 at 300 K suggests that Mu$_{\rm p}$ is ionized with a small activation energy ($\sim10$ meV). Such characteristics lead to the speculation that Mu$_{\rm p}$ is a shallow-level state whose neutral state is stable only at low temperatures. 

In summary, the majority of implanted Mu in a-IGZO:H forms a diamagnetic state that may correspond to Mu$^-$-H$^-$ complex at V$_{\rm O}$, while a small probability of falling into a shallow-donor state remains, depending on the distance between Mu and H.  

%\section*{ACKOWLEDGMENTS}
We thank TRIUMF and J-PARC staff for their technical support during muon experiment. Thanks are also to A. Suter, Z. Salman, and T. Prokschar for the technical support and fruitful discussion during Low Energy Muon experiment at PSI. We also acknowledge CROSS-Tokai for providing facilities for sample-handling in their user laboratories. This work was financially supported by the MEXT Elements Strategy Initiative to Form Core Research Center at Tokyo Institute of Technology. The DFT calculation was performed at KEK supercomputer under the project numbers 16/17-18, 15/16-07, 14/15-13, 13/14-09, and T12/13-02.

%\newpage


\begin{thebibliography}{99}%\label{sec:TeXbooks}%
\bibitem{Hosono:06} H. Hosono, J. Non-Cryst. Solids {\bf 203}, 334 (1996); {\bf 352}, 851 (2006).

\bibitem{Kamiya:10} T. Kamiya and H. Hosono, NPG Asia Mater. {\bf 2}, 15 (2010).

\bibitem{Robertson:12} J. Robertson, J. Non-Cryst. Solids 358, 2437 (2012).

\bibitem{Ghaffarzadeh:10} K. Ghaffarzadeh, A. Nathan, J. Robertson, S. Kim, S. Jeong, C. Kim, U. Chung, and J. Lee, Appl. Phys. Lett. {\bf 97}, 113504 (2010); {\bf 97}, 143510 (2010).

\bibitem{Lee:10} D. H. Lee, K. Kawamura, K. Nomura, T. Kamiya, and H. Hosono, Electrochem. Solid State Lett. {\bf 13}, H324 (2010).

\bibitem{Chowdhury:10} M. D. H. Chowdhury, P. Migliorato, and J. Jang,  Appl. Phys. Lett. {\bf 97}, 173506 (2010).

\bibitem{Park:12} J. S. Park, W. J. Maeng, H.-S. Kim, and J.-S. Park, Thin Solid Films {\bf 520}, 1679 (2012).

\bibitem{Jeon:12} S. Jeon, S. E. Ahn, I. Song, C. J. Kim, U. I. Chung, E. H. Lee, I. Yoo, A. Nathan, S. Lee, J. Robertson, and K. Kim, Nature Mat. {\bf 11}, 301 (2012).

\bibitem{Jeong:13} J. K. Jeong, J. Mater. Res. {\bf 28}, 2071 (2013).

%\bibitem{Nomura:08} K. Nomura, T. Kamiya, E. Yanagi, E. Ikenaga, K. Yang, K. Kobayashi, M. Hirano, and H. Hosono, Appl. Phys. Lett. {\bf 92}, 202117 (2008).

\bibitem{Nomura:11} K. Nomura, T. Kamiya, E. Ikenaga, H. Yanagi, K. Kobayashi, and H. Hosono, J. Appl. Phys. {\bf 109}, 073726 (2011).

\bibitem{Nomura:19} K. Ide, K. Nomura, T. Kamiya, and H. Hosono, Phys. Status Solidi A {\bf 216}, 1800372 (2019).

\bibitem{KamiyaJDT09} T. Kamiya, K. Nomura and H. Hosono, J. Display Technol. {\bf 5}, 273 (2009), and references therein.

\bibitem{BangAPL17} J.~Bang, S.~Matsuishi and H.~Hosono,  Appl. Phys. Lett. {\bf 110}, 232105 (2017).

\bibitem{Li:17}  H. Li, Y. Guo, and J. Robertson,  Sci. Rep. {\bf 7}, 16858 (2017); Phys. Rev. Mater. {\bf 2}, 074601 (2018).

%\bibitem{Li:18}  H. Li, Y. Guo, and J. Robertson,  

\bibitem{Okabe:18} H. Okabe, M. Hiraishi, S. Takeshita, A. Koda, K. M. Kojima, and R. Kadono,
Phys. Rev. B. 98 (2018) 075210(1-6).

\bibitem{TangSSST17}  H. Tang, Y. Kishida, K. Ide, Y. Toda, H. Hiramatsu, S. Matsuishi, S. Ueda, N. Ohashi, H. Kumomi, H. Hosono, and T. Kamiya,  ECS J. Solid State Sci. Technol. {\bf 6} P365 (2017).

\bibitem{Holzschuh:82} E. Holzschuh, W. K\"undig, P. F. Meier, B. D. Patterson, J. P. F. Sellschop, M. C. Stemmet, and H. Appel, Phys. Rev. A {\bf 25}, 1272 (1982).

\bibitem{Kiefl:86} R. F. Kiefl, W. Odermatt, Hp. Baumeler, J. Felber, H. Keller, W. K\"undig, P. F. Meier, B. D. Patterson, J. W. Schneider, K. W. Blazey, T. L. Estle, and C. Schwab, Phys. Rev. B {\bf 34}, 1474 (1986). 

\bibitem{HayanoPRB79} R. S.~Hayano, Y. J. Uemura, J. Imazato, N. Nishida, T. Yamazaki, and R. Kubo,  Phys. Rev. B {\bf 20}, 850 (1979).

\bibitem{KojimaJPSConf17} K. M.~Kojima, T. Murakami, Y. Takahashi, H. Lee, S. Y. Suzuki, A. Koda, I. Yamauchi, M. Miyazaki, M. Hiraishi, H. Okabe, S. Takeshita, R. Kadono, T. Ito, W. Higemoto, S. Kanda, Y. Fukao, N. Saito, M. Saito, M. Ikeno, T. Uchida, and M. M. Tanaka,  J. Phys. Conf. Ser.  {\bf 551}, 012063, (2014).

\bibitem{trim:91} W. Eckstein, {\it Computer simulation of Ion-Solid Interactions} (Springer, Berlin, 1991).

\bibitem{KojimaPRB04} K. M. Kojima, J. Yamanobe, H. Eisaki, S. Uchida, Y. Fudamoto, I. M. Gat, M. I. Larkin, A. Savici, Y. J. Uemura, P. P. Kyriakou, M. T. Rovers, and G. M. Luke,  Phys. Rev. B {\bf 70}, 094402 (2004). 

\bibitem{CoxJPhys06} S.F.J. Cox, J. L. Gavartin, J. S. Lord, S. P. Cottrell, J. M. Gil, H. V. Alberto, J. Piroto Duarte, R. C. Vilao, N. Ayres de Campos, D. J. Keeble, E. A. Davis, M. Charlton, and D. P. van der Werf,  J. Phys.: Condens. Matter. {\bf 18}, 1079 (2006).

\bibitem{VASP} G. Kresse and J. Hafner,  Phys. Rev. B {\bf 47} , 558 (1993); {\bf 49}, 14251 (1994). 

\bibitem{CGvWalle:01} C. G. Van de Walle,  Phys. Rev. Lett. {\bf 85}, 1012 (2001).

\bibitem{ShimomuraPRL02} K. Shimomura, K. Nishiyama and R. Kadono,  Phys. Rev. Lett. {\bf 89}, 255505 (2002).

\bibitem{Kilic:02} \c C. Kili\c c and A. Zunger, Appl. Phys. Lett. {\bf 81}, 73 (2002).

\bibitem{Crook:97} M. R. Crook and R. Cywinski, J. Phys.: Condens. Matter {\bf 9}, 1149 (1997).

\bibitem{Kadono:08} R. Kadono, K. Shimomura, K. H. Satoh, S. Takeshita, A. Koda, K. Nishiyama, E. Akiba, R. M. Ayabe, M. Kuba, and C. M. Jensen, Phys. Rev. Lett. {\bf 100}, 026401 (2008).

\bibitem{Beck:75} R. Beck, P. F. Meier, and A. Schenck, Z. Physik B {\bf 22}. 109 (1975).

%\bibitem{CoxRevProgPhys09} S.F.J~Cox, {\it Rep. Prog. Phys.} {\bf 72}  116501 (2009).


\end{thebibliography}
\end{document}